\documentclass[12pt]{article}
\input epsf
\usepackage{epsfig}
\def\bC{{\mathbf{\overline{C}}}}
\def\cross{{\tt{x}}}
\def\C{{\mathbf{C}}}
\def\o{{\tt{o}}}
\def\x{{\tt{x}}}
\def\Z{{\mathbf{Z}}}
\def\R{{\mathbf{R}}}
\def\mod{{\mathrm{mod}\ }}
\def\deg{{\mathrm{deg}\ }}
\def\const{{\mathrm{const}}}
\def\id{{\mathrm{id}}}
\def\Arg{{\mathrm{Arg}\ }}
\begin{document}
\title{Zeros of eigenfunctions of some anharmonic oscillators}
\author{Alexandre Eremenko\footnote{Supported by NSF grants
DMS-0555279 and DMS-0244547.}, $\;$ Andrei Gabrielov and
Boris Shapiro}
\maketitle
\begin{center}
{\bf 1. Introduction}
\end{center}
\vspace{.2in}

We consider eigenvalue problems of the form
\begin{equation}
\label{problem}
-y^{\prime\prime}+P(z)y=\lambda y,\quad y(-\infty)=y(\infty)=0,
\end{equation}
where $P$ is a real even
polynomial with positive leading coefficient, which 
is called a potential. 
The boundary
condition is equivalent to 
$y\in L^2(\R)$ in this case.
It is well-known that the spectrum is discrete,
and all eigenvalues
$\lambda$ are real and simple, see, for example \cite{BS,S}.
The spectrum
can be arranged in an increasing
sequence $\lambda_0<\lambda_1<\ldots$. 

Eigenfunctions $y$ are real entire functions
of order $(\deg P+2)/2$ and each of them
has finitely many real zeros. The number of real zeros of an
eigenfunction is equal to the subscript of the corresponding
eigenvalue $\lambda_k$. Asymptotic behavior
of complex zeros of eigenfunctions is well-known,
in particular, their arguments
accumulate to finitely many directions,
the so-called Stokes' directions \cite{Bank,Bank2}.
Using this one can show that for a real even potential $P$ of degree $4$
with positive leading coefficient,
{\em all but finitely many} zeros of each eigenfunction
lie on the imaginary axis. See also \cite{Trinh} where a similar
result was obtained for some cubic potentials.

\vspace{.1in}

\noindent
{\bf Theorem 1.} {\em Let $P$ be a real even polynomial
of degree $4$ with positive leading coefficient.
Then all non-real zeros of
eigenfunctions $y$ of the problem $(\ref{problem})$
belong to the imaginary axis.}
\vspace{.1in}

Under the assumptions of Theorem 1, every eigenfunction
has infinitely many imaginary zeros.

For the special case that $P(z)=cz^4+d,\; c>0, d\in\R$
Theorem~1 was conjectured by Titchmarsh \cite[p. 147]{Tit}
and proved by Hille \cite[p. 617-618]{Hille}, see also
\cite[p. 188-190]{Hille2}.

Operator considered in Theorem~1 is called a quartic
anharmonic oscillator
in quantum mechanics, and it was studied extensively
by physicists and mathematicians.
A very brief survey of known results is contained in Chapter I
of \cite{Ushveridze}.

For some potentials
$P$, there exist eigenfunctions
with finitely many zeros. Such eigenfunctions have the form 
\begin{equation}
\label{eig}
y(z)=Q(z)\exp T(z),
\end{equation}
with polynomials $Q$ and $T$. For example, this is the case
when $P$ is of degree $2$; then {\em all} eigenfunctions
are of the form (\ref{eig}), and $Q$'s are the Hermite
polynomials. Eigenfunctions with finitely many zeros
can also
occur for polynomials $P$
of higher degree, and these situations are of interest
to physicists \cite{Lopez,Shifman,Turbiner,Ushveridze}.  
It is easy to see that eigenfunctions of the form (\ref{eig})
can exist only in the case that 
$\deg P\equiv 2\,(\mod 4)$. Moreover,
for every $n$ such that $n\equiv 2\,
(\mod 4)$ there exist real polynomials $P$ such that some
eigenfunctions have the form (\ref{eig}), see \cite{EM}.
\vspace{.1in}

\noindent
{\bf Theorem 2.} {\em Let $P$ be a
real even polynomial of degree $6$
with positive leading coefficient.
If $(\ref{eig})$ is
an eigenfunction of $(\ref{problem})$,
then all non-real zeros of $Q$ belong to the imaginary
axis.}
\vspace{.1in}

Since the union of the real and imaginary axes does not
contain any Stokes directions
for a sextic potential $P$ (see Section 2),
we conclude that eigenfunctions of a sextic
potential with infinitely many zeros
cannot have all zeros in the union of the real and imaginary axes.

In the proof of Theorem 2 we obtain a classification
of eigenfunctions (\ref{eig}) which can occur in
operators (\ref{problem}) with even sextic potential.
It turns out that this classification fits the
classification of the so-called
``quasi-exactly solvable'' sextic potentials
\cite{Lopez,Ushveridze}. As a corollary we obtain in
Section~6 that
for even sextic potentials,
eigenfunctions (\ref{eig}) can
occur
only for Lie-algebraic
quasi-exactly solvable sextic potentials listed in
\cite{Lopez,Ushveridze}. More precisely,
Theorem~2 combined with the results of
Turbiner and Ushveridze \cite{Ushveridze} gives the
following
\vspace{.1in}

\noindent
{\bf Corollary.} {\em Let $P$ be a real even polynomial
of degree $6$, and suppose that problem $(\ref{problem})$
has at least one solution $y$ of the form $(\ref{eig})$.
Then
$$P(z)=c^2z^6+2bcz^4+\{ b^2-c(4m+2p+3)\} z^2+\const,$$
where $c\in\R\backslash\{0\},\; b\in\R,\; p\in\{0,1\}$
and $m$ a non-negative integer.}
\vspace{.1in}

It was shown by Turbiner and Ushveridze that these potentials
have exactly $m+1$ linearly independent
eigenfunctions of the form (\ref{eig}). They correspond
to the first $m+1$ even numbered eigenvalues if $p=0$
and to the first $m+1$ odd-numbered eigenvalues if $p=1$.

The proofs of theorems 1 and 2 are of purely topological nature,
they are based on the study of the action of the symmetry
group $\Z_2\times\Z_2$ of the problem (\ref{problem})
on certain partitions
of the complex $z$-plane associated with
the eigenfunctions. 

Phenomenon described in Theorems 1 and 2 occurs only
for potentials of degrees $4$ and $6$:
\vspace{.1in}

\noindent
{\bf Theorem 3.} {\em For every $k\geq 2$,
there exists a real even polynomial $P$
of degree $4k+2$ with positive leading coefficient, such that 
the problem $(\ref{problem})$ has 
an eigenfunction of the form $(\ref{eig})$,
but the zero set of $Q$ 
is not a subset
of the union of the real and imaginary axis.}
\vspace{.1in}

Polynomial $P$ in Theorem~3 does not belong
to the classification of quasi-exactly solvable potentials
that arise from finite-dimensional
Lie algebras of differential operators in \cite{Lopez}. 

The third-named author is sincerely grateful to
A.~Turbiner for the hospitality at UNAM in October 2006
and inspiring discussions on the location of the roots
of eigenfunctions of Schr\"odinger operators.
\vspace{.2in}

\noindent
\begin{center}
{\bf 2. Preliminaries}
\end{center}
\vspace{.2in}

From now on, we always assume that $P$ is real, even
and has positive leading coefficient. We denote $d=\deg P$. 

Making the change of the independent
variable $z\mapsto-z$ we conclude that
every eigenfunction is either even or odd. We normalize
even eigenfunctions by the condition 
$$y(0)=1,$$
and the odd ones by the condition
$$y'(0)=1.$$
Consider another solution $y_1$ of the differential equation
in (\ref{problem})
normalized by
$$y_1(0)=0,\quad y_1^\prime(0)=1$$
in the case that the eigenfunction $y$ is even,
and 
$$y_1(0)=1,\quad y_1^\prime(0)=0$$
in the case that the eigenfunction $y$ is odd.
Then $y_1$ is even or odd, and its parity is opposite
to the parity of $y$. 
Thus the meromorphic function $f=y/y_1$ is real and odd,
in particular it is symmetric with respect to both
real and imaginary axes: if we denote the reflections
with respect to the coordinate axes by 
$$R(z)=\overline{z}\quad\mbox{and}\quad I(z)=-\overline{z},$$
then
\begin{equation}
\label{symmetry}
f\circ R=R\circ f\quad\mbox{and}\quad f\circ I=I\circ f.
\end{equation}
The following facts are well-known \cite{Nev1,Nev2,S}.
The rays
$$\rho_j=\left\{ t\exp\{\pi i(2j-1)/(d+2)\}:
\; 0<t<\infty\right\},\quad 0\leq j\leq d+1$$
are the Stokes' directions. They divide
the plane into $d+2$ sectors $S_j$,
where $S_j$ is bounded by
$\rho_{j}$ and $\rho_{j+1}$.
In each sector, each non-zero solution of the differential equation
in (\ref{problem}) 
exponentially tends either to $0$ or to $\infty$,
(on every ray from the origin in this sector),
in particular,
$y(z)\to 0$ in $S_0$ and $S_{d/2+1}$ in view of the boundary conditions
in (\ref{problem}), 
while $y_1$ tends to $\infty$ in these two sectors.

Notice that the set of zeros and the set of poles
of $f$ are both invariant with respect to $R$ and $I$.

Meromorphic function $f$ is of order $(d+2)/2$,
has no critical points (which means that $f'(z)\neq 0$ and
all poles are simple) and
the set of its asymptotic values is finite.
Such functions have been studied in great detail in
\cite{EM,GO,Nev0,Nev1,Nev2,S,W}. If $A$ is the set of asymptotic values, then
the restriction
$$f:\C\backslash f^{-1}(A)\to\bC\backslash A$$
is an (unramified) covering, and also $f$ is unramified at
preimages of $A$.

In each sector $S_j$, the function
$f$ tends to an asymptotic value $a_j$ exponentially, and the
asymptotic values in adjacent sectors are distinct.
More precisely, for every sufficiently small $\epsilon>0$,
$$f(re^{i\theta})\to a_j,\quad r\to\infty,$$
uniformly with respect to $\theta\in[\rho_j+\epsilon,
\rho_{j+1}-\epsilon].$
We have $a_0=a_{d/2+1}=0$,
and the symmetry properties (\ref{symmetry})
imply that $a_j=R(a_{-j})$, and $a_j=I(a_{d/2+1-j})$. 
Here we understand the index $j$ as a residue modulo $d+2$.

The only singularities of the inverse function $f^{-1}$
are logarithmic branch points; they all lie over the
asymptotic values. The total number of the logarithmic
branch points is $d+2$, and they correspond to the $d+2$ sectors~$S_j$.

As an example, consider the situation in Theorem~1,
where $d+2=6$. If we denote $a=a_1$ then the symmetry
relations (\ref{symmetry}) imply that the asymptotic
values are $$(a_0,a_1,a_2,a_3,a_4,a_5)=(0,a,I(a),0,-a,R(a)).$$
Now, the condition that $a_1\neq a_2$ implies that $a$ cannot
belong to the imaginary axis. We will later see in the course
of the proof of Theorem~1 that $a$ cannot be real.

In Theorem~2, we have $d+2=8$ and the form of the eigenfunction
(\ref{eig}) shows that $a_{2k}=0,\; k=0,1,2,3$.
Denoting $a_1=a$ again, we obtain from the symmetry relations
(\ref{symmetry}) that 
$$(a_1,a_3,a_5,a_7)=(a,I(a),-a,R(a)).$$
We will see in the course of the proof of Theorem~2 that
$a$ can be neither real nor pure imaginary.
We conclude that in both theorems 1 and 2 $f$ has five 
asymptotic values, 
\begin{equation}\label{asvalues}
a,I(a),-a,R(a)\quad\mbox{and}\quad 0.
\end{equation} 

To study topological properties of the function $f$ one considers
the pullback by $f$ of an appropriate cell decomposition of the
Riemann sphere. The usual choice of this cell decomposition
leads to an object which is called the line complex  
\cite{Drape,EM,GO,Nev0,Nev1,Nev2,W}.
However classical line complexes
are not convenient for our purposes because they do not
reflect the symmetry relations (\ref{symmetry}), see, for example
\cite{EM}. So in the main part of the proofs
of theorems 1 and 2 (Sections 3 and 4)
we use slightly different approach.

However we find it more convenient to use the standard line complexes
in the proof of Theorem 3. So we recall the definition
of a line complex in the beginning of Section~5.

\vspace{.2in}

\begin{center}
\nopagebreak{\bf 3. Common part of the proofs of Theorems 1 and 2}
\end{center}
\nopagebreak
\vspace{.2in}

In this Section,
$f$ is a meromorphic function of finite order,
with no critical points and five asymptotic values
as in (\ref{asvalues}),
where $a$ is neither real nor imaginary,
and satisfies the symmetry
conditions (\ref{symmetry}).
We will treat the simpler case of real
or imaginary $a$ separately, in the end of Section~4.

We will work with partitions of a topological space $X$ which can
be either the plane
$\C$ or the Riemann
sphere $\bC$ into subsets which we call vertices, edges and faces.
All our partitions are locally finite, that is every point in $X$ has
a neighborhood that intersects only finitely many edges,
faces and vertices.

A {\em vertex} is just a point in $X$.
An {\em edge} in $X$ is the image of the interval $(0,1)$
under a continuous map $\phi:[0,1]\to X$ whose restriction on
$(0,1)$ is injective. 
The points $\phi(0)$ and $\phi(1)$ are the endpoints of the edge
(they do not belong to the edge but always belong to $X$). The endpoints may
be equal.
We also say that the edge connects
$\phi(0)$ with $\phi(1)$. The degree of a vertex $x$ of a partition
is defined as the total number of ends of edges whose endpoints coincide
with this vertex. Thus an edge with $x\in\{\phi(0),\phi(1)\}$,
may contribute one or two units to the degree of $x$,
one if $\phi(0)\neq\phi(1)$ and two
if $\phi(0)=\phi(1)$.
A {\em face} is a simply connected domain in $X$
whose boundary is locally connected.

A {\em partition} is a representation of $X$ 
as a locally finite
disjoint union of faces,
edges and vertices, such that all endpoints
of all edges are vertices, 
and the boundary of
every face consists of edges and vertices.
We {\em do not} require that the closure of a face be homeomorphic
to a closed disc.

We begin with the partition
$C$ of the Riemann sphere $\bC$ which
consists of:

\begin{itemize}
\item
One vertex, $\infty$,
\item
Four edges $L_k,\; 1\leq k\leq 4$, each beginning and ending
at $\infty$, and such that $L_k$ separates one of the
asymptotic values $a,I(a),-a,R(a)$ from all
other asymptotic values. Moreover,
we require that the union of these edges be invariant with
respect to $R$ and $I$, more precisely, $L_2=I(L_1),\; L_3=-L_1$
and $L_4=R(L_1)$. 
\item
Five faces $D_k,\; 0\leq k\leq 4$ which are
the components of 
$\C\backslash\displaystyle\cup_{k=1}^4L_k$. 
We enumerate them so that $0\in D_0$, and $\partial D_k=L_k\cup\{\infty\},\;
1\leq k\leq 4$.
\end{itemize}
\vspace{.1in}

This partition is shown in Figure 1. In this illustration,
$a$ belongs to the first quadrant. 
\begin{center}
\epsfxsize=2.5in%
\centerline{\epsffile{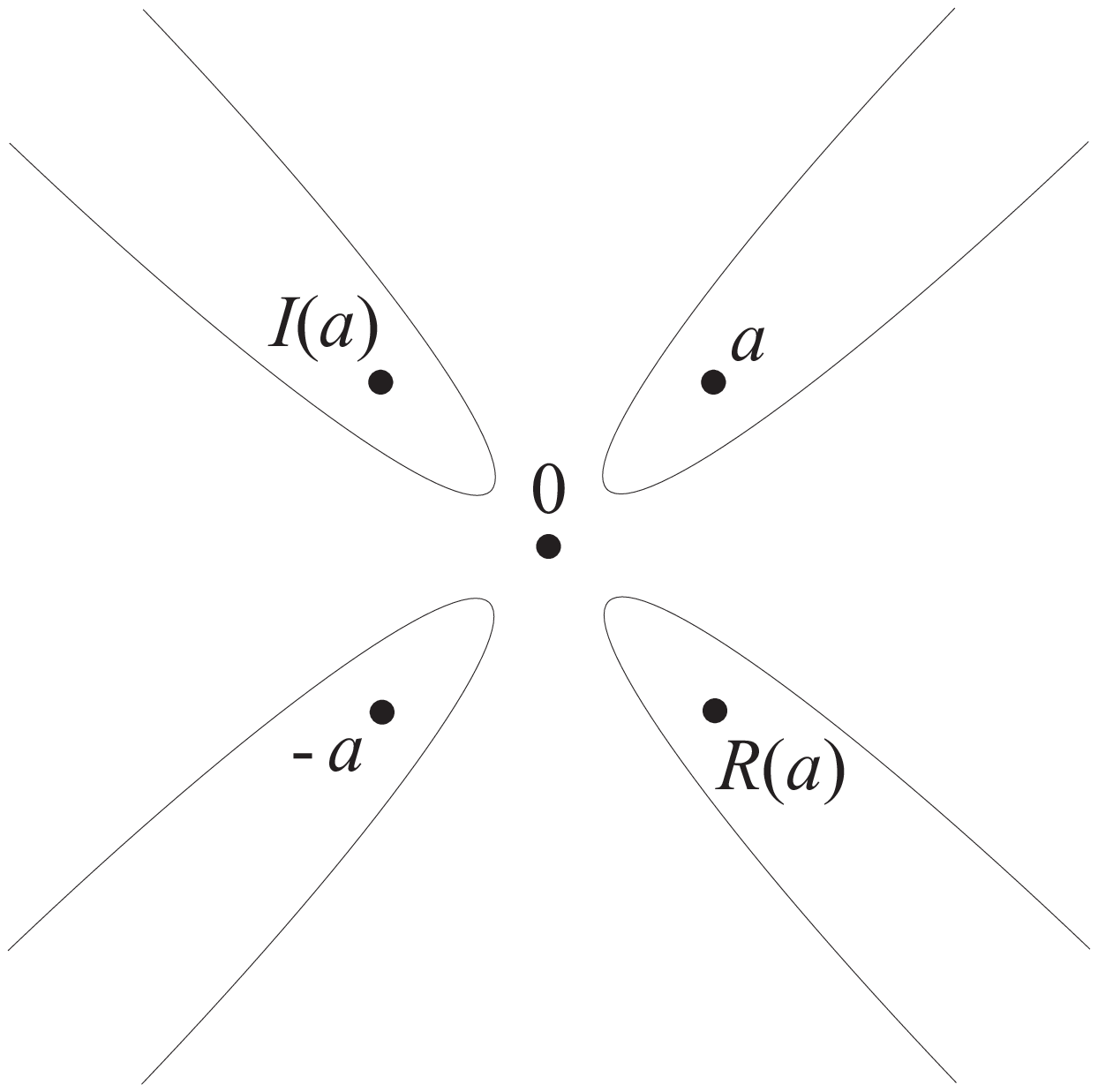}}
\nopagebreak
Fig. 1. Partition $C$ of the Riemann sphere.
\end{center}
\vspace{.1in}

Now we consider the preimage of $C$ under $f$.
This is a partition $\Phi$ of the plane. 
Faces, edges and vertices of $\Phi$ are defined
as 
components of the $f$-preimages of faces, edges and vertices
of $C$.

So the vertices of $\Phi$ coincide with the poles of $f$.
The degree of each vertex
is $8$, the same as the degree of the vertex in the partition $C$.

The edges of $\Phi$ are disjoint curves connecting vertices.
The edges can be of two types: an edge connecting a vertex to itself
is called a {\em loop}, and an edge connecting two distinct vertices
is called a {\em simple edge}. 

We classify the faces of $\Phi$ into three types: 
\vspace{.1in}

\noindent
(i) Bounded faces whose boundaries consist of a loop and a vertex.
We call them {\em loop-faces}. The closure of each loop-face is
mapped by $f$ homeomorphically onto the closure of one
of the $D_k$ with $1\leq k\leq 4$. 
\vspace{.1in}

\noindent
(ii) Bounded faces which are mapped
by $f$ homeomorphically
onto $D_0$. (This homeomorphism does not necessarily
extend to a homeomorphism between the closures!)  Each of these faces contains
exactly one zero of $f$, and each zero of $f$ belongs to
one of these faces. We call them {\em zero-faces}. 

\vspace{.1in}

\noindent
(iii) Unbounded faces. The restriction of $f$ onto
an unbounded face is a universal covering over $D_k\backslash b_k$
for some $k\in\{0,\ldots,4\}$. Here $b_k$ is the asymptotic
value which is contained in $D_k$.
The boundary of each unbounded face consists of countably many edges
and countably many vertices.
\vspace{.1in}

\noindent
We label all faces by the corresponding asymptotic values:
if a face is a component of the preimage of $D_k$ then its
label is the asymptotic value $b_k\in D_k$.
Bounded faces labeled by $b_k$ with $1\leq k\leq 4$
are the loop-faces, while bounded faces labeled by $0$
are the zero-faces. 

We will need the following properties of the partition
$\Phi$.
\vspace{.1in}

\noindent
1. If two faces have a common boundary edge then their
labels are distinct.
\vspace{.1in}

\noindent
2. The $1$-skeleton of $\Phi$ is connected. Indeed,
the loops $L_k,\; 1\leq k\leq 4$
generate the fundamental group of
$\bC\backslash\{ b_0,\ldots,b_4\}$. 
As
$$f:\C\backslash f^{-1}(\{ b_0,\ldots,b_4\})\to
\bC\backslash \{ b_0,\ldots,b_4\}$$
is a covering, we conclude that every pair of
poles of $f$ can be connected by a curve which belongs to
the $1$-skeleton of $\Phi$.
\vspace{.1in}

\noindent
3. Every edge connecting two different vertices belongs to the
boundary of some unbounded face. Indeed, suppose that an edge $e$
connecting two different vertices belongs to the boundaries
of two bounded faces $F_1$ and $F_2$. As the labels
of these two faces are distinct (by property 1 above),
one of these labels is not $0$.
But a bounded face whose
label is not $0$ has to be a loop-face (see (i)).
So $e$ has to be a loop, which contradicts the assumption.
\vspace{.1in}


Now we transform our partition $\Phi$ into a tree.
This is done in two steps.
\vspace{.1in}

\noindent
{\em Step 1.} Remove all loop edges and all loop-faces. The resulting 
partition of the plane is called $\Phi'$. Each face $F'$ of $\Phi'$
is a union of a face $F$ of $\Phi$ with some loops and loop-faces of $\Phi$.
We label
$F'$ in $\Phi'$ by the same label as $F$ had in $\Phi$.
So all bounded faces of $\Phi'$
are now labeled by $0$. They are Jordan regions with
at least two boundary edges and at least two boundary vertices. 
It easily follows from the property 2 of $\Phi$
that $\Phi'$ has connected $1$-skeleton.
Moreover, $\Phi'$ is invariant
with respect to both $I$ and $R$.
\vspace{.1in}

\noindent
{\em Step 2.} Every bounded zero-face $F'$ of $\Phi'$ contains a unique
zero of $f$.
We call this zero an $\o$-vertex and define a new partition $\Phi^{\prime\prime}$
of the plane in the following way.
The vertices of the new partition are the vertices
of $\Phi'$, which we call now $\cross$-vertices,
and the new vertices which are called $\o$-vertices.
In other words, $\o$-vertices are the zeros of $f$
and $\cross$-vertices are the poles of $f$.
To define the edges of $\Phi^{\prime\prime}$
we connect each $\o$-vertex in a zero-face $F'$ of $\Phi'$
to each $\cross$-vertex on the boundary of $F'$ by a new edge inside $F'$,
so that these new edges are disjoint. This is possible to do because
the closure of $F'$ is locally connected.
Then we remove all edges of
$\Phi'$ on the boundary of $F'$. We perform this operation
on every zero-face $F'$ of $\Phi'$.
\vspace{.1in}

Let us show that on Step 2,
we can choose the new edges of $\Phi^{\prime\prime}$
in such a way that $\Phi^{\prime\prime}$ is symmetric
with respect to both $R$ and $I$.
Indeed, there are three possibilities
for the orbit
of a zero-face $F'$ of $\Phi'$ under the action of
the group $\Z_2\times\Z_2$ generated by $I$ and $R$:
the orbit of $F'$ 
can consist
of one, two or four faces.
We consider these possibilities separately.

If the orbit consists of $4$ elements
then $F'$ is neither $R$- nor $I$-invariant. 
We choose the edges connecting the $\o$-vertex
in $F'$ to the boundary $\cross$-vertices arbitrarily
(with the only condition that they are disjoint), and then
in other faces of the orbit of $F'$ we
use the images of these edges under the action of the group.

Suppose now
that the orbit of $F'$ consists of two elements,
for example, $F'$ is $R$-invariant but not $I$-invariant. 
We first define the new edges in $F'$, so that the union
of these new edges is $R$-invariant.
This can be done if we notice
that an $R$-invariant simply connected region intersects the
real line by an interval.  If some endpoint $x$ of this interval is
a $\cross$-vertex, we connect $x$ to $\o$ by an interval of
the real line. If $x\in\partial F'$ is a 
$\cross$-vertex in the upper half-plane, we connect it with $\o$ by
a curve in the intersection
of $F'$ with the upper half-plane, so that these curves
for different $\cross$-vertices are disjoint.
(We use here the fact that an intersection of a
simply connected $R$-symmetric region
with the upper half-plane is always connected,
and its boundary is locally
connected if the boundary of $F'$ is.)
Finally if $x\in\partial F'$ is in the lower half-plane,
we use the $R$-image of the edge connecting $\o$ with $R(x)$.
Then we define the new edges in the other face $I(F')$ of
the orbit of $F'$ as the $I$-images of the edges in $F'$.
The procedure for an $I$-invariant but not $R$-invariant
face is the same.
There are no other possibilities for an orbit of two
elements:
if $F'=-F'$ then $F'$ has to be invariant with respect to
the whole group, since $F'$ is simply connected and 
a centrally symmetric simply connected region
has to contain $0$.

The remaining case of $F'$ which is both $R$- and $I$- invariant is 
treated similarly. The intersection of such face with the coordinate
cross $\R\cup i\R$ consists of the union of two symmetric intervals,
one on the real axis, another on the imaginary axis.
To the $\cross$-vertices
at the endpoints of these intervals (if there are any
such vertices) we draw straight edges from $\o$.
Then we notice
that the intersection of $F'$ with the first quadrant is
connected and has
locally connected closure. So we can draw the edges
from $\o$ to the $\cross$-vertices in the first quadrant so that
these edges are contained in the first quadrant. The remaining edges in $F'$
are the images of those in the first quadrant under the symmetry group action.
\vspace{.1in}

The following Proposition summarizes the needed properties of
$\Phi^{\prime\prime}$
\vspace{.1in}

\noindent
{\bf Proposition 1}.
{\em The partition $\Phi^{\prime\prime}$ has the following properties.

\noindent
a) Its $1$-skeleton is an infinite tree properly embedded in the plane. 

\noindent
b) Every edge belongs to the boundaries of two faces
with distinct labels.

\noindent
c) An $\o$-vertex cannot
belong to the boundary of a face labeled $0$.

\noindent
d) Each $\cross$-vertex is either connected
with an $\o$-vertex by an edge or belongs to the
boundary of a face labeled $0$.

\noindent
e) $\Phi^{\prime\prime}$ has $d+2$ ends and $d+2$ faces.

\noindent
f) Faces $F_0$ and $F_{d/2+1}$ in a
counter-clockwise order have labels $0$
and these faces are interchanged by $I$.} 
\vspace{.1in}

{\em Proof.}
First we prove that
$\Phi^{\prime\prime}$
has no bounded faces.
Suppose that $F'$ is
a bounded face of $\Phi'$. It is a Jordan
region with some number $k\geq 2$ boundary edges (see the description
of Step 2).
On Step $2$ we replaced $F'$ by certain number of triangles (=regions
bounded by three edges),
one triangle for each boundary edge of $F'$, and then glued this
triangle to some face $F^\prime_0$ of $\Phi'$,
exterior to $F'$, along this edge.
This face $F^\prime_0$ was unbounded by property 3 above.
So it remains unbounded
after adding triangles along some edges. So step 2
destroys all
bounded faces and does not create new ones. Thus the
$1$-skeleton of $\Phi^{\prime\prime}$ is a forest.
That it is connected,
follows from connectedness of the $1$-skeleton of $\Phi'$;
evidently Step 2 does not destroy connectedness.

It follows that the $1$-skeleton of $\Phi^{\prime\prime}$ is
a tree.  
As the vertices and edges
accumulate only to infinity, the tree is properly embedded in $\C$. This proves~a).

To prove b), we first notice that each edge of $\Phi^{\prime\prime}$
that comes from $\Phi'$ belongs to the boundaries of
two faces with distinct labels (property 1 above).
So it remains to prove b)
for the new edges added on Step 2.
Let $e$ be an edge added on step $2$. It connects
an $\o$ and $\x$
inside a zero-face $F'$ of $\Phi'$.
Let $e_1$ and $e_2$ be the two edges on the boundary of $F'$ incident to
the common vertex $\x$.
Then $e_1$ and $e_2$ are $f$-preimages of two
different edges $L_k$ and $L_j$ of $C$,
and thus the faces $F_1^\prime$
and $F_2^\prime$ exterior
to $F'$ that have $e_1$ and $e_2$ on their
boundaries have distinct labels. Thus $e$ is a common boundary edge
of two faces with distinct labels. This proves b).

To prove c), we consider an $\o$-vertex and
the bounded face $F'$ of $\Phi'$
which contains this vertex. This face $F'$ is labeled by $0$
(as all bounded faces of $\Phi'$ are),
 and thus there cannot be
an edge in $\Phi'$ in the common boundary of $F'$ and
another face labeled by $0$.
It follows that Step 2 
cannot produce an $\o$-vertex on the boundary of
a face with label $0$.   

To show d), we first notice that
the number of unbounded faces
and the labeling of these faces do not change when we
perform steps 1 and 2.
Now statement e), as well as f) follow
from the asymptotic properties of $f$ stated in
the Preliminaries:
each unbounded face (of any partition $\Phi$, $\Phi'$ or
$\Phi^{\prime\prime}$)
is asymptotic to one of the sectors $S_j$
where $f(z)$ tends to the label of this face. 
The total number of sectors is $d+2$, and the
bisectors of two of them are the positive and negative 
rays of the real line.

This completes the proof of Proposition 1.
\vspace{.1in}

In the next Section we will classify all
trees satisfying the
conclusions of the Proposition 1 with $d=4$.
For $d=6$ we will use an additional property
that arises from (\ref{eig}): that every other
face of $\Phi^{\prime\prime}$ is labeled by $0$.

It will result from this
classification that all trees
arising from the eigenfunctions in
theorems 1 and 2 as $1$-skeletons of $\Phi^{\prime\prime}$
have the following property:
{\em each $\o$-vertex is
either fixed by $R$ or fixed by $I$.}
As the $\o$-vertices are the
zeros of $f$ and thus the zeros of the
eigenfunction $y$, this will prove our theorems 1 and 2. 
\vspace{.1in}

\begin{center}
{\bf 4. Completion of the proofs of theorems 1 and 2}
\end{center}
\vspace{.1in}

We classify the embedded trees up to homeomorphisms of the plane that
commute with both $R$ and $I$,
send vertices to vertices  and
preserve the labels of faces and vertices
($\o$ and $\cross$).
\vspace{.1in}

\noindent
{\bf Proposition 2.} {\em For $d=4$, there are only three 
types of embedded trees satisfying the conclusions of Proposition~1;
they are shown in Fig. 2.
All $\o$-vertices lie in the union of the real and
imaginary lines.  
All trees of these three types are parametrized by a non-negative 
integer $n$: the number of $\o$-vertices on the real line. First type occurs
when $n=0$, second for even $n\geq 2$ and third when $n$ is odd.}
\vspace{.1in}
\begin{center}
\epsfxsize=5.0in%
\centerline{\epsffile{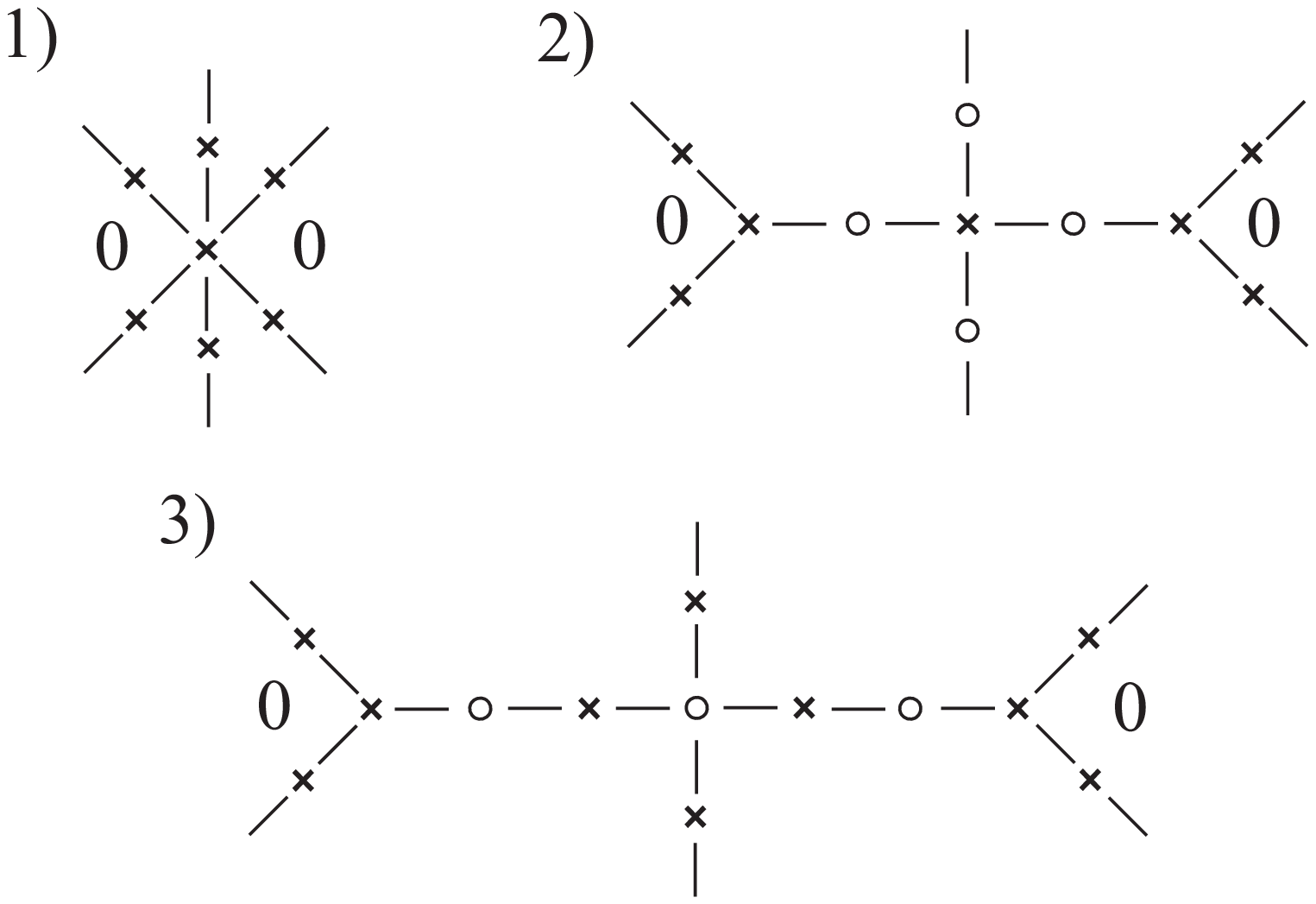}}
\nopagebreak
Fig. 2. Types of possible trees in Theorem 1.
\end{center}
\vspace{.1in}

{\em Proof.} If vertices of order $2$ are ignored,
there are two topological types of properly embedded
trees
with $6$ ends, both $I$ and $R$ symmetries and satisfying
b) and f) of Proposition~1:
\vspace{.1in}
\begin{center}
\epsfxsize=4.0in%
\centerline{\epsffile{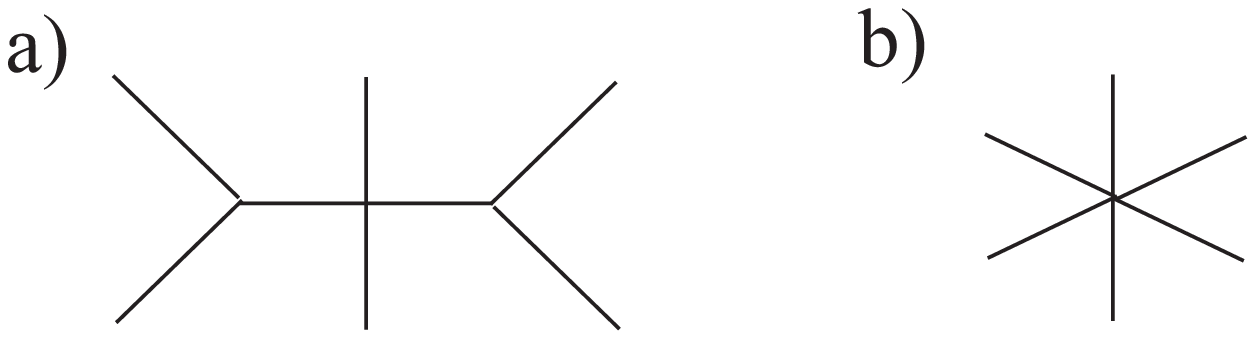}}
\nopagebreak
Fig. 3. Topological types of possible trees in Theorem 1 (ignoring
vertices of order $2$).
\end{center}
\vspace{.1in}
This simple fact can be proved along the same
lines as Proposition A in the Appendix and it is left to
the reader. Now statements c) and d) of Proposition~1 imply that
the $\o$-vertices lie on the coordinate cross.
\vspace{.1in}

For Theorem 2, we have
one property in addition to those stated
in Proposition 1:
every even-numbered face (in the natural cyclic order) is
labeled by~$0$.
\vspace{.1in}

\noindent
{\bf Proposition 3.} {\em For $d=6$,
there are only five types of embedded
trees satisfying the
conclusions of Proposition 1 and the additional property
that every even-numbered face is labeled by $0$. They are shown
in Figure 4.
These trees are parametrized by two integers: the total number $m$
of $\o$-vertices
and the number $n$ of $\o$-vertices on the real line. These integers
satisfy the following evident restrictions:
$0\leq n\leq m$ and $n-m$ is even.}
\vspace{.1in}
\begin{center}
\epsfxsize=5.0in%
\centerline{\epsffile{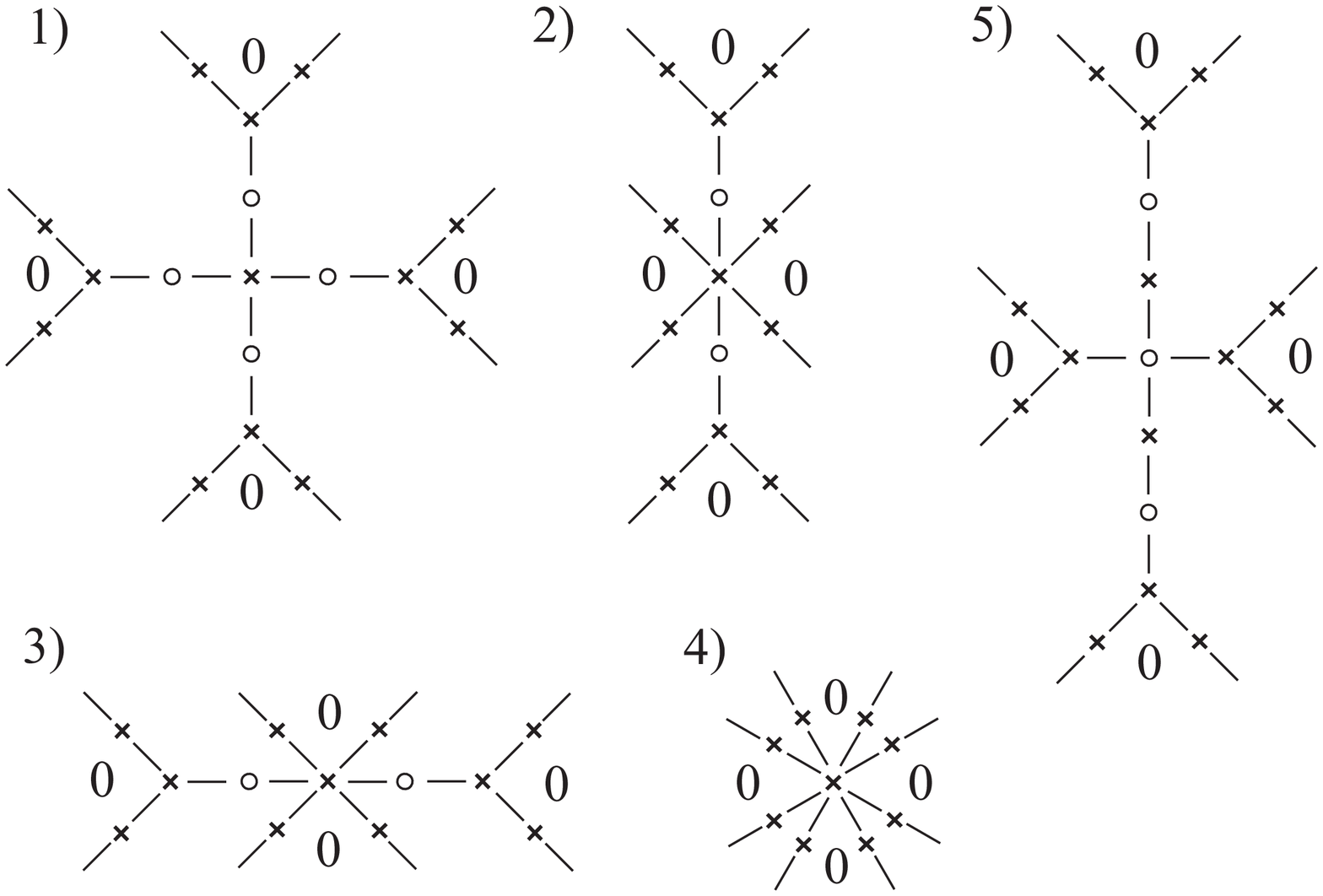}}
\nopagebreak
Fig. 4. Possible types of trees in Theorem 2.
\end{center}
\vspace{.1in}

{\em Proof.} Only topological types a), c), f) and j)
of Proposition A in the Appendix satisfy conditions b)
and f) of Proposition 1. Topological type a) gives two
types of trees, depending on the type of vertex in the
center.
\vspace{.1in}

One can show, using a result of Nevanlinna cited in the next Section
that all types of
trees described in propositions 2 and 3 can actually occur
for the eigenfunctions in theorems 1 and 2.
\vspace{.1in}

To complete the proof of Theorem~1, it remains to prove that $a$
cannot be real.
The proof is by contradiction.
If $a$ is real, we have only three asymptotic values, $0,a,-a$.
We repeat with simplifications the construction in 
Section~3. The partition $C$ of the Riemann sphere has two
edges and three faces now:
\vspace{.1in}
\begin{center}
\epsfxsize=2.5in%
\centerline{\epsffile{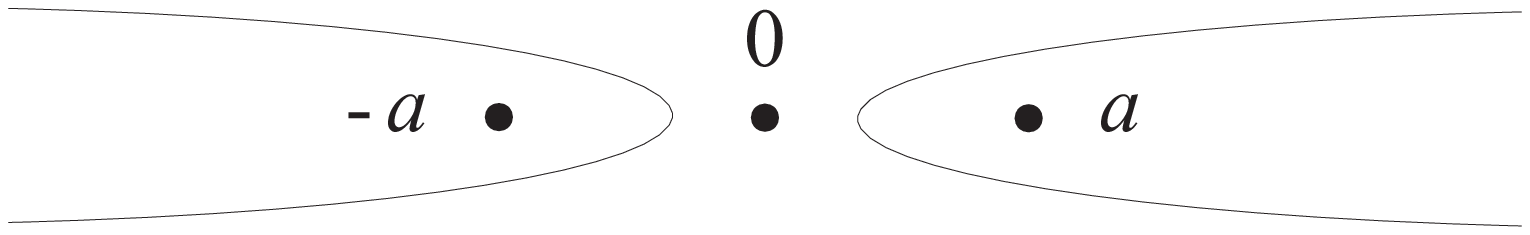}}
\nopagebreak
Fig. 5. Partition $C$ in the case of real $a$.
\end{center}
\vspace{.1in}

The degree of each vertex of $\Phi$ is $4$.
The boundary of each zero-face in $\Phi$ contains two edges and two vertices,
and thus it contains no loops. Proposition 1 remains true for the
partition $\Phi^{\prime\prime}$.
But now an analog of the Propositions 2 and 3
shows that no
admissible graphs exist. Indeed, graphs of the types 2 and 3 in Fig. 2
are
excluded because two faces labeled $a$ have common edges in them.
Graph of the type 1 is excluded because it has a vertex of
degree 6. For the same reasons, all graphs in Fig. 4 are excluded.
Thus $a$ cannot be real.

To complete the proof of Theorem~2, we have to show that $a$ can
be neither real nor imaginary. The argument is the same as we just used
to show that $a$ cannot be real in Theorem~1, and it is left to the reader. 

\vspace{.2in}
\begin{center}
{\bf 5. Theorem 3}
\end{center}
\vspace{.2in}

We begin by recalling the definition of a line complex.
Let $f$ be a meromorphic function of finite order
without
critical points and with finitely many asymptotic values.

Let $\Gamma$ be an
oriented Jordan curve in the Riemann sphere $\bC$,
which passes
once through each asymptotic value.
Orientation of $\Gamma$ induces a cyclic order on
the set of asymptotic values. We denote the asymptotic values
in this order by
$(b_1,\ldots,b_q)$.
The subscripts $1,\ldots,q$ in $b_k$
are considered as residues modulo $q$,
so that $b_{q+1}=b_1$, etc.
The curve $\Gamma$ with marked points $b_k$
is called the {\em base curve},
and the points $b_k$ are called the {\em base points}.

Choose one point $\o$ inside $\Gamma$ and one point $\cross$
outside $\Gamma$. Then connect $\o$ and $\cross$ by $q$ 
disjoint edges
$\Gamma_1,
\ldots,\Gamma_q$, so that each
$\Gamma_k$
intersects $\Gamma$ exactly once, and this
intersection happens
on the open arc $(b_k,b_{k+1})$ of $\Gamma$.

The points $\o$ and $\cross$, the curves $\Gamma_k$
and the components of the complement to the union of these
curves and points form a partition of the
Riemann sphere with two vertices, $q$ edges and $q$ faces.

The $f$-preimage of this partition  
of the plane $\C$
is traditionally called the {\em line complex} $L$.
The faces, edges and vertices
of the line complex can be naturally {\em labeled}
by the names of their images. 
 A line complex is 
considered as a topological object: two line complexes are
equivalent if they can be mapped one onto another by
an orientation preserving homeomorphism of the plane.

The following properties of a line complex are known
\cite{Drape,GO,Nev0,Nev1,Nev2} and easy to prove.

The $1$-skeleton of $L$ is a bi-partite graph
(in particular it has no loops).
Every vertex belongs to the boundaries of exactly $q$
faces, and the labels of these $q$ faces are
all distinct;
the labels $(a_1,\ldots,a_q)$ of these faces
follow anti-clockwise around an $\o$-vertex and clockwise
around a $\cross$-vertex.

The faces are of two types: $2$-gons and $\infty$-gons.
The $2$-gons are mapped by $f$ homeomorphically onto
neighborhoods of the base points. So each $2$-gon
contains exactly one simple
solution of the equation $f(z)=b_k$,
where $b_k$ is the label of this $2$-gon.
$\infty$-gons correspond to the logarithmic branch points
of $f^{-1}$, and the restriction of $f$ onto an $\infty$-gon
is a universal cover of a punctured neighborhood of the
base point $b_k$,
where $b_k$ is the label of this $\infty$-gon.
In particular, the equation
$f(z)=b_k$ has no solutions in an $\infty$-gon labeled $b_k$.

An unlabeled
line complex can be
defined intrinsically, without any reference
to $f$. It is a partition of the plane into vertices, edges and faces
with the following property:
\vspace{.1in}

\noindent
{\em The $1$-skeleton is a connected bi-partite properly embedded graph
whose all vertices have the same degree $q$.}
\vspace{.1in}

If such partition of the plane is given, there are two choices
of labeling the vertices with $\cross$ and $\o$.
Furthermore, if a cyclically ordered set
of points $(b_1,\ldots,b_q)$ is given, we can always label the faces
of our partition with elements of this set, so that the subscripts
of the labels increase counter-clockwise around  each $\o$-vertex
and clockwise around a $\cross$-vertex. Notice that such labeling
is uniquely defined once the label of one face is specified,
and the label of one face can be prescribed arbitrarily.

A fundamental theorem which is due to Nevanlinna \cite{Nev1},
see also \cite{GO},
says the following.
\vspace{.1in}

{\em Suppose that a labeled line complex is given with $d+2$ unbounded faces
and all bounded faces are $2$-gons. Choose a base curve passing through
the labels according to their cyclic order.
Then there exists a meromorphic function $f$ in the plane of order $(d+1)/2$
with no critical points and whose
line complex with respect to
the this base curve is equivalent to the given one. 

This function is unique up to a change of the independent variable $z\mapsto
cz+b,\; c\neq 0$.}
\vspace{.1in}

Each $f$ given by this theorem is a ratio of two linearly independent
solutions of the differential equation $-y^{\prime\prime}+Py=0$
where 
$$-2P=\frac{f^{\prime\prime\prime}}{f'}-
\frac{3}{2}\left(\frac{f^{\prime\prime}}{f'}
\right)^2.$$

We use this result to prove Theorem 3.
For simplicity of illustrations,
we consider only the case  $d=10$.
Then we should have $12$ sectors $S_j$, and we choose the asymptotic
values in these sectors to be $a_0=a_2=a_4=a_6=a_8=a_{10}=0$
and
$$(a_1,a_3,a_5,a_7,a_9,a_{11})=(a,b,a,R(a),R(b),R(a)),$$
where $a$ and $b$ are distinct and  belong to the positive imaginary axis,
for example one can take $a=2i$ and $b=i$ as we do in our pictures.
So we have $5$ asymptotic values.
Then we choose the imaginary axis (oriented ``up'') as our base curve,
the $\o$ point at $-1$ and the $\cross$ at $+1$.

Now we consider the following line complex:
\begin{center}
\epsfxsize=3.0in%
\centerline{\epsffile{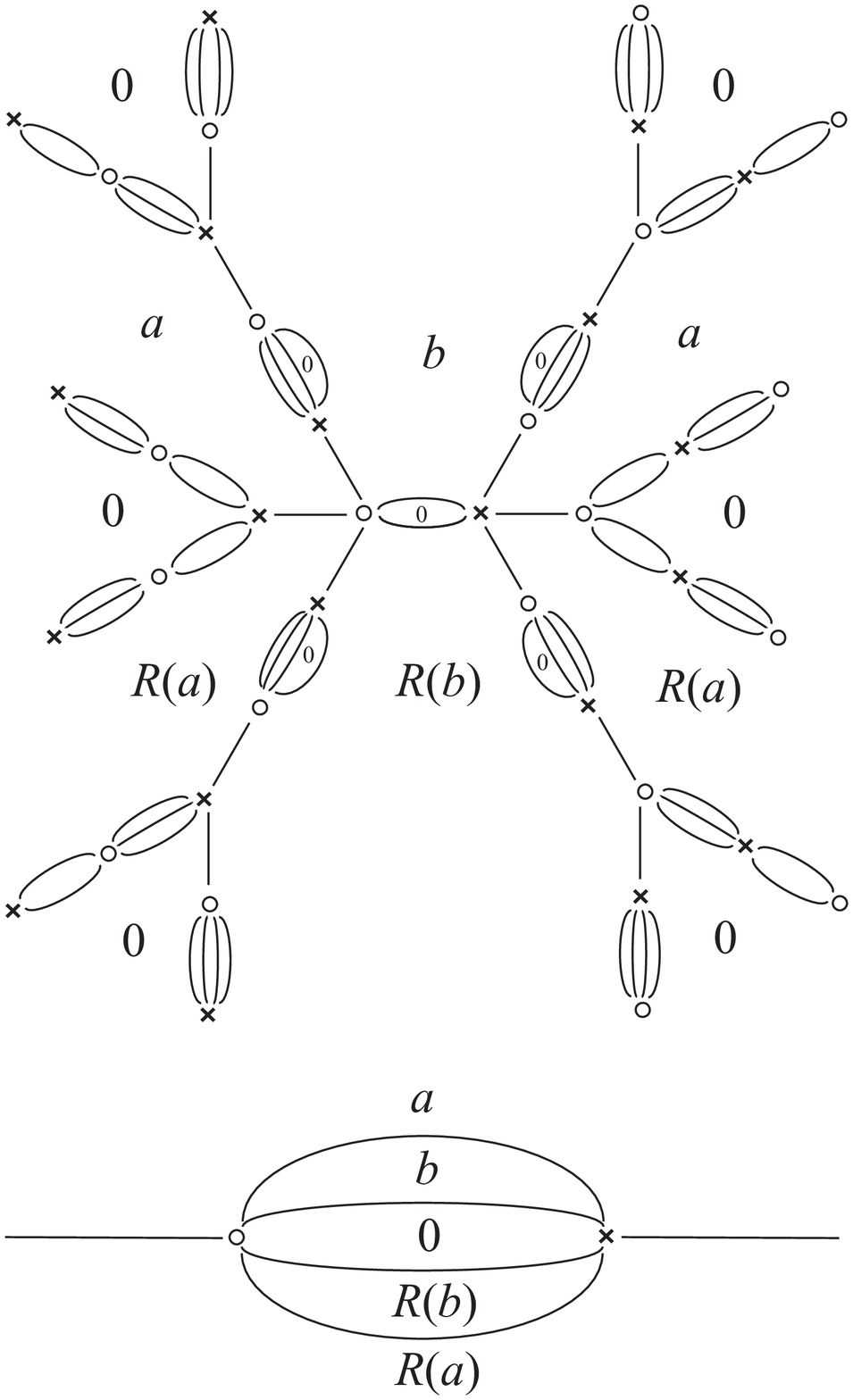}}
\nopagebreak
Fig. 6. Line complex for Theorem 3. Curves $\Gamma_k$ are
shown in the bottom.
\end{center}
\vspace{.1in}

Meromorphic function $f$ corresponding to this complex by
Nevanlinna's theorem will be normalized as follows:
$\o^*=-1$ and $\x^*=1$, where $\o^*$ and $\x^*$ are the
two vertices closest to the center of the
picture. Then it is clear that $f$ has
the necessary symmetry properties
(\ref{symmetry}). Four zeros shown in the picture form
a single orbit under the $\Z_2\times\Z_2$ action.
So these zeros are neither real nor imaginary. 
\vspace{.2in}

\begin{center}
{\bf 6. Quasi-exactly solvable sextic potentials}
\end{center}
Here we prove the Corollary stated in the Introduction.

In 1987, Turbiner and Ushveridze (see, a detailed exposition in
\cite{Ushveridze}) discovered the family
of sextic potentials
$$P_{m,p,b}(z)=z^6+2bz^4+(b^2-4m-2p-3)z^2,$$
where $m$ is a non-negative integer, $p\in\{0,1\}$,
and $b\in\R$,
with the following remarkable property.
Problem (\ref{problem}) with $P=P_{m,p,b}$
has $m+1$ eigenfunctions
of the form
$$y(z)=
z^pQ_{k,m,p,b}(z^2)
\exp\left(\frac{z^4}{4}-b\frac{z^2}{2}\right),\quad 0\leq k
\leq m$$ 
where $Q_{k,m,p,b}$ are polynomials of degree $m$.
These eigenfunctions correspond to the eigenvalues
$\lambda_{2k+p}$
of the Schr\"odinger operators (\ref{problem}) with
$P=P_{m,p,b}$. Such potentials, having several
first eigenfunctions of the form (\ref{eig}),
are sometimes called quasi-exactly solvable
\cite{Ushveridze}.

Using the method which goes back to Stieltjes 
\cite{St},
Ushveridze proved that all zeros of the
polynomials $Q_{k,m,p,b}$
are real \cite{Ushveridze}, pp. 53--56,
so all zeros of the eigenfunctions belong
to the union of the real and imaginary axes.
More precisely,
$Q_{k,m,p,b}$ has $k$ positive and $m-k$
negative zeros. 

On the other hand, in the proof of our Theorem 2 we obtained
the classification of {\em all}
meromorphic functions $f=f_{k,n,p,a}$ arising
from eigenfunctions of the form (\ref{eig})
for all problems (\ref{problem})
with even polynomials $P$ of degree $6$. 
Here $f_{k,n,p,a}$ is the function $f$ from Section~2
with $2n+p$ zeros, $2k+p$ of them real, and asymptotic
value $a$ in the first quadrant.

Proportional functions $f$ belong to the same potential
$P$ and the same eigenvalue,
hence we can normalize so that $a=\exp(i\alpha)$
where $\alpha \in (0,\pi/2)$ when $p=1$
and $\alpha\in(-\pi/2,0)$ when $p=0$.

Thus we obtain a family of
functions $\alpha=g_{ k,m,p}(b)$. To each potential
$P_{m,p,b}$ and each $k\in[0,m]$ this function $g$ puts
into correspondence the argument of the
asymptotic value $a$ of the
corresponding function
$f=y/y_1$, where $y$ is the $2k+p$-th eigenfunction
of (\ref{problem}), and
$y_1$ is a second linearly independent
solution of the differential equation in (\ref{problem}),
with $P=P_{m,p,b}$ and $\lambda=\lambda_{2k+p}$. This function $f$ is
$f_{k,m,p,a}$ for some $a$, and we define
$g_{k,m,p}(b)=\Arg a$. 
Moreover, $f$ is normalized so that $|a|=1$ and
$0<|\Arg a|<\pi/2$.

Thus each function $g$ maps the real line into an
open interval $J\in\R$,
where
$J=(0,\pi/2)$ for $p=1$, and 
$J=(-\pi/2,0)$ for $p=0$.
We have to prove that all these functions $g$ are
surjective.

It is well-known that $g$ is continuous and real analytic
(see, for example, \cite{S}).
On the other hand, for each $a\in J$, there is a
neighborhood $V\subset J$ of $a$, such that a (right)
inverse
branch $h:J\to\R$  of $g$ can be defined, 
that is $h\circ\phi=\id_J$, and moreover,
this $h$ is real analytic \cite{S}.
We conclude that $g$ is
surjective and this proves the Corollary. 
\vspace{.1in} 

\begin{center}
{\bf Appendix. Classification of trees}
\end{center}
\vspace{.2in}

In this Appendix we classify all trees with $8$ ends, that
have no ends on the real and imaginary axes,
and are both $R$- and $I$-symmetric.
In what follows we call such trees {\em double-symmetric}.
We say that two double-symmetric
trees are isomorphic if there
exists  an orientation-preserving and
commuting with both $R$
and $I$ homeomorphism
of the complex plane sending one to the other. We do
not require that this homeomorphism sends vertices to
vertices.
\vspace{.1in}

\noindent
{\bf Proposition A}
{\em There exist $11$ non-isomorphic
double-symmetric trees with no ends
fixed by either $R$ or $I$, see Figure~7.}
\vspace{.1in}
\begin{center}
\epsfxsize=5.0in%
\centerline{\epsffile{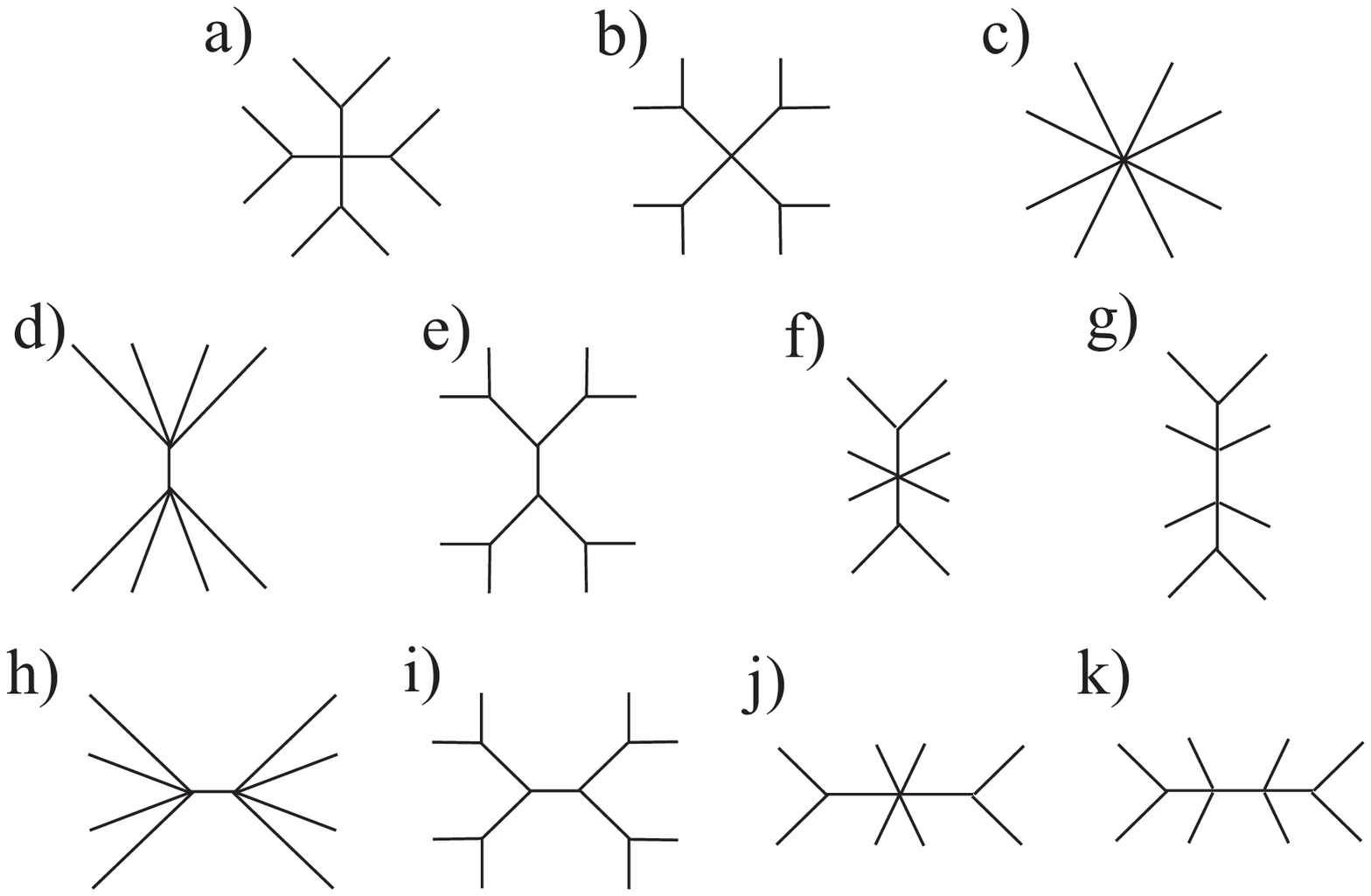}}
\nopagebreak
Fig. 7. Double symmetric trees with $8$ ends
none of which is on the axes.
\end{center}
\vspace{.1in}

The proof uses the following
simple lemma classifying symmetric planar
rooted trees with 4 ends whose proof is left
to the reader.
\vspace{.1in}

\noindent
{\bf Lemma.}
{\em There exist 6 non-isomorphic planar rooted trees
with 4 ends (not counting the root)
and no vertices of degree $2$,
symmetric with respect to a reflection in a line passing through the root,
see Figure~8.}
\begin{center}
\epsfxsize=3.0in%
\centerline{\epsffile{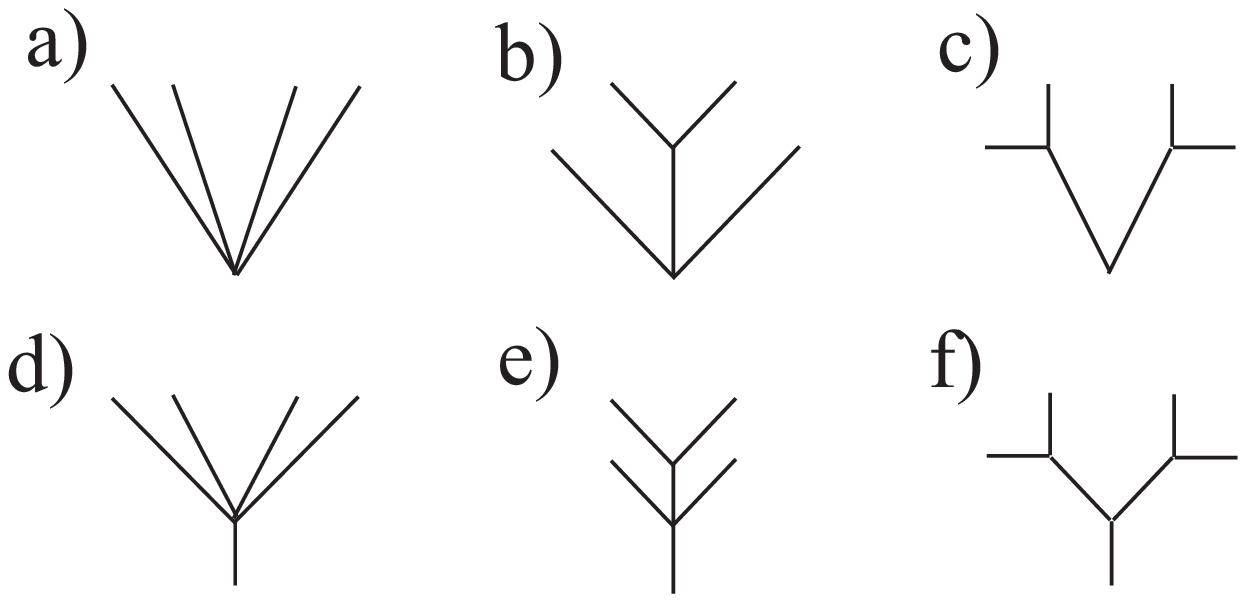}}
\nopagebreak
Fig. 8. $I$-symmetric rooted trees with $4$ ends.
\end{center}
\vspace{.1in}

To prove Proposition~A we argue as follows.
Any double-symmetric tree contains the origin
(since a doubly symmetric $1$-complex
embedded in the plane and not containing 
the origin either is either
disconnected or contains a cycle).
Consider the intersection of such a tree with the coordinate axes.
If it intersects both axes non-trivially, i.e., not only at the origin,
then the only possibility is the tree marked $a$ in Figure~7.
Assume now that a double-symmetric tree intersects the real axis only at
the origin. Then the intersection of that tree
with the closed upper half-plane is a planar tree rooted at the origin
with 4 ends, symmetric with respect to $I$.
By the Lemma, there are exactly  6 such trees.
They correspond to the trees marked $b$ through $g$ in Figure~7.
To get all the remaining double-symmetric trees we have to rotate the
latter $6$ trees by $\pi/2$. Notice that trees marked $b$ and $c$ are invariant
under this rotation, producing just $4$ new trees, namely, trees
marked  $h$ through $k$ in Figure~7.
Thus the total number of trees equals $1+6+4=11$.\hfill

{\em Purdue University

West Lafayette, IN

47907--2067 U.S.A.

eremenko@math.purdue.edu

agabriel@math.purdue.edu
\vspace{.2in}

Stockholm University

Stockholm, S-10691, Sweden

shapiro@math.su.se}
\end{document}